\documentclass[twocolumn,aps,showpacs,prl,superscriptaddress]{revtex4} 

\usepackage{sidecap}
\usepackage{ulem}
\usepackage{epsfig}
\usepackage{amsmath,amssymb,amsthm}
\usepackage{graphicx}
\usepackage{bm}
\usepackage{color}

\DeclareMathAlphabet{\mathpzc}{OT1}{pzc}{m}{it}

\begin{document}

\renewcommand{\textfraction}{0.00}


\newcommand{\vAi}{{\cal A}_{i_1\cdots i_n}} 
\newcommand{\vAim}{{\cal A}_{i_1\cdots i_{n-1}}} 
\newcommand{\vAbi}{\bar{\cal A}^{i_1\cdots i_n}}
\newcommand{\vAbim}{\bar{\cal A}^{i_1\cdots i_{n-1}}}
\newcommand{\htS}{\hat{S}} 
\newcommand{\htR}{\hat{R}}
\newcommand{\htB}{\hat{B}} 
\newcommand{\htD}{\hat{D}}
\newcommand{\htV}{\hat{V}} 
\newcommand{\cT}{{\cal T}} 
\newcommand{\cM}{{\cal M}} 
\newcommand{\cMs}{{\cal M}^*}
\newcommand{\vk}{\vec{\mathbf{k}}}
\newcommand{\bk}{\bm{k}}
\newcommand{\kt}{\bm{k}_\perp}
\newcommand{\kp}{k_\perp}
\newcommand{\km}{k_\mathrm{max}}
\newcommand{\vl}{\vec{\mathbf{l}}}
\newcommand{\bl}{\bm{l}}
\newcommand{\bK}{\bm{K}} 
\newcommand{\bb}{\bm{b}} 
\newcommand{\qm}{q_\mathrm{max}}
\newcommand{\vp}{\vec{\mathbf{p}}}
\newcommand{\bp}{\bm{p}} 
\newcommand{\vq}{\vec{\mathbf{q}}}
\newcommand{\bq}{\bm{q}} 
\newcommand{\qt}{\bm{q}_\perp}
\newcommand{\qp}{q_\perp}
\newcommand{\bQ}{\bm{Q}}
\newcommand{\vx}{\vec{\mathbf{x}}}
\newcommand{\bx}{\bm{x}}
\newcommand{\tr}{{{\rm Tr\,}}} 
\newcommand{\bc}{\textcolor{blue}}

\newcommand{\beq}{\begin{equation}}
\newcommand{\eeq}[1]{\label{#1} \end{equation}} 
\newcommand{\ee}{\end{equation}}
\newcommand{\bea}{\begin{eqnarray}} 
\newcommand{\eea}{\end{eqnarray}}
\newcommand{\beqar}{\begin{eqnarray}} 
\newcommand{\eeqar}[1]{\label{#1}\end{eqnarray}}
 
\newcommand{\half}{{\textstyle\frac{1}{2}}} 
\newcommand{\ben}{\begin{enumerate}} 
\newcommand{\een}{\end{enumerate}}
\newcommand{\bit}{\begin{itemize}} 
\newcommand{\eit}{\end{itemize}}
\newcommand{\ec}{\end{center}}
\newcommand{\bra}[1]{\langle {#1}|}
\newcommand{\ket}[1]{|{#1}\rangle}
\newcommand{\norm}[2]{\langle{#1}|{#2}\rangle}
\newcommand{\brac}[3]{\langle{#1}|{#2}|{#3}\rangle} 
\newcommand{\hilb}{{\cal H}} 
\newcommand{\pleft}{\stackrel{\leftarrow}{\partial}}
\newcommand{\pright}{\stackrel{\rightarrow}{\partial}}

\title{Heavy flavor puzzle at LHC: \\
 a serendipitous interplay of jet suppression and fragmentation}

\date{\today}
 
\author{Magdalena Djordjevic}
\affiliation{Institute of Physics Belgrade, University of Belgrade, Serbia}

\begin{abstract} 
Both charged hadrons and D mesons are considered to be excellent probes of QCD matter created in ultra relativistic heavy ion collisions. Surprisingly, recent experimental observations at LHC show the same jet suppression for these two probes, which - contrary to pQCD expectations -  may suggest similar energy losses for light quarks and gluons in QCD medium. We here use our recently developed energy loss formalism in a finite size dynamical QCD medium to analyze this phenomenon that we denote as the ``heavy flavor puzzle at LHC''. We show that this puzzle is a consequence of an unusual combination of the suppression and fragmentation patterns, and in fact does not require invoking the same energy loss for light partons.  Furthermore, we show that this combination leads to a simple relationship between the suppressions of light hadrons and D mesons and the corresponding bare quark suppressions. Consequently, a coincidental matching of jet suppression and fragmentation allows considerably simplifying the interpretation of the corresponding experimental data.
\end{abstract}


\maketitle

\section{Introduction}
Jet suppression~\cite{Bjorken} is considered an excellent probe of QCD matter created in ultra-relativistic heavy ion collisions~\cite{Gyulassy,DBLecture,Wiedemann2013,Vitev2010}. Specifically, charged hadrons and D mesons are often in the focus of both theoretical and experimental research, since they represent the most direct probes of light and heavy flavor in QCD matter. However, while D meson suppression is indeed a clear charm quark probe, this is not the case for light hadrons. That is, while experimentally measured prompt D mesons are exclusively composed of charm quarks~\cite{ALICE_preliminary}, light hadrons are composed of both light quarks and gluons~\cite{Vitev0912}. Furthermore, gluons are known to have larger jet energy loss compared to light and heavy quarks, while light and charm quarks are expected to have similar suppressions~\cite{MD_PRC}. It is therefore clearly expected that charged hadron suppression should be significantly larger compared to D meson suppression. However, preliminary ALICE measurements~\cite{ALICE_h,ALICE_preliminary} surprisingly show that charged hadrons and D mesons have the same $R_{AA}$. We here denote this surprising observations as the ``heavy flavor puzzle at LHC'', and explaining this puzzle is the main goal of this paper. 

We note that, perhaps the most straightforward explanation of the puzzle would be that, contrary to pQCD expectations, gluons and light quarks in fact lose the same amount of energy in the medium created at ultra relativistic heavy ion collisions. In fact, the same possibility was also suspected in the context of RHIC experiments~\cite{Gyulassy_viewpoint,SE_puzzle}, where similar suppressions were observed for pions and single electrons; this has led some theorists to seek explanations outside conventional QCD~\cite{FNG,NGT,GC,HGB}. The main goal of this paper is to analyze phenomena behind the ``heavy flavor puzzle at LHC'', and investigate if pQCD is able to explain such unexpected experimental data.

\section{Results and discussion} 
To analyze the puzzle described in the previous section, we will here use our recently developed theoretical formalism, which is outlined in detail in~\cite{MD_LHSupp_2013}. The procedure is based on {\it i)} jet energy loss in a finite size dynamical QCD medium~\cite{MD_PRC,DH_PRL,MD_PRC2012,MD_Coll}, {\it ii)} finite magnetic mass effects~\cite{MD_MagnMass}, {\it iii)} running coupling~\cite{MD_LHSupp_2013}, {\it iv)} multigluon fluctuations~\cite{GLV_suppress}, {\it v)} path-length fluctuations~\cite{WHDG,Dainese} and {\it v)} most up to date production~\cite{Cacciari:2012,Vitev0912} and fragmentation functions~\cite{DSS}.  Note that the computational procedure uses no free parameters, i.e. the parameters that are stated below correspond to the standard literature values. 

We consider a QGP of temperature $T{\,=\,}304$\,MeV (as extracted by ALICE~\cite{Wilde2012}), with 
$n_f{\,=\,}2.5$ effective light quark flavors. Perturbative QCD scale is taken to be $\Lambda_{QCD}=0.2$~GeV. For the light quarks we assume 
that their mass is 
dominated by the thermal mass $M{\,=\,}\mu_E/\sqrt{6}$, and the gluon mass is  $m_g=\mu_E/\sqrt{2}$~\cite{DG_TM}. Here Debye mass $\mu_E \approx 0.9$~GeV is obtained by self-consistently solving  Eq.~(3) from~\cite{MD_LHSupp_2013} (see also~\cite{Peshier2006}), and magnetic mass $\mu_M$ is taken as $0.4 \, \mu_E < \mu_M < 0.6 \, \mu_E$~\cite{Maezawa,Bak}. Charm and bottom mass is, respectively, $M=1.2$~GeV and $M=4.75$~GeV. For charm and bottom, the initial quark 
spectrum, $E_i d^3\sigma(Q)/dp_i^3$, is computed at next-to-leading order
using the code from~\cite{Cacciari:2012,FONLL}; for gluons and light quarks, 
the initial distributions are computed at next-to-leading order as 
in~\cite{Vitev0912}. For light hadrons, we use DSS fragmentation 
functions~\cite{DSS}. For D mesons we use BCFY fragmentation functions~\cite{BCFY}. Path length distributions are extracted from~\cite{Dainese}, while fragmentation functions are implemented according to~\cite{Vitev06}.

\begin{figure}
\epsfig{file=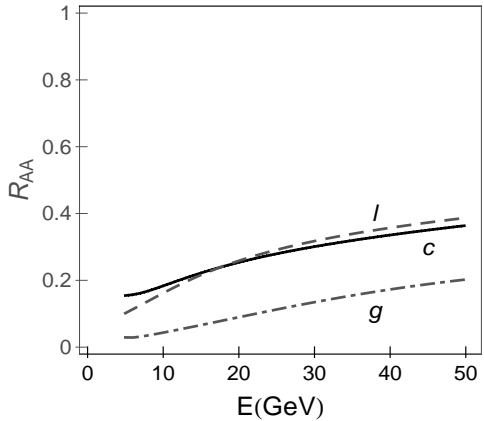,width=2.5in,height=2.2in,clip=5,angle=0}
\vspace*{-0.3cm}
\caption{{\bf Suppression of quark and gluon jets.}  Momentum dependence of the jet suppression is shown for charm quarks (the full curve), light quarks (the dashed curve) and gluons (the dot-dashed curve). 
Electric to magnetic mass ratio is fixed to $\mu_M/\mu_E =0.5$.}
\label{QuarkELossRaa}
\end{figure}

\begin{figure}
\epsfig{file=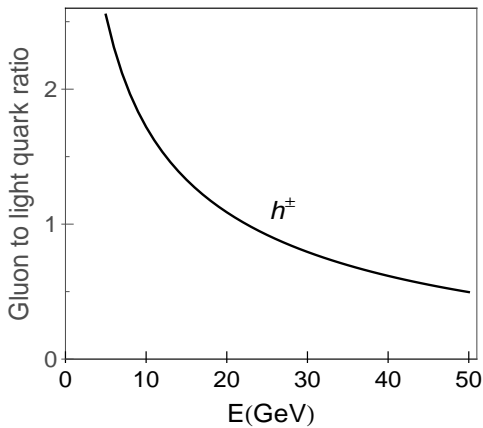,width=2.5in,height=2.2in,clip=5,angle=0}
\vspace*{-0.3cm}
\caption{{\bf Ratio of gluon to light quark contribution  in the initial distributions of charged hadrons  as a function of momentum. } The figure shows the ratio of the gluon to light quark contribution in the initial distributions of charged hadrons.}
\label{GLRatio}
\end{figure}

We start by quantitatively reproducing the expectations summarized in the introduction, in order to obtain a clear view of the relevant hierarchies  for the suppression and the initial distributions. Fig.~\ref{QuarkELossRaa} shows the comparison of the suppressions for light and charm quark and gluon jets. We see that suppression of gluon jets is significantly larger compared to corresponding suppression of quark jets, while the suppression predictions for light and charm quarks are similar. Furthermore, in Fig.~\ref{GLRatio} we show that both light quarks and gluons significantly contribute to the charged hadron production - in fact, in the lower momentum range, gluons dominate over light quarks; therefore both contributions from gluons and light quarks have to be taken into account when analyzing charged hadron suppression. On the other hand, D mesons present a clear charm quark probe, since the feedown from B mesons is subtracted from the experimental data~\cite{ALICE_preliminary}.

\begin{figure*}
\epsfig{file=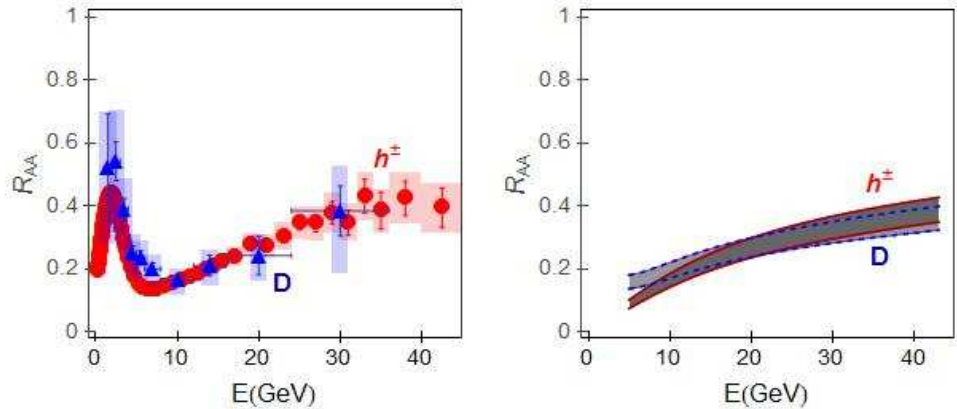,width=5in,height=2.3in,clip=5,angle=0}
\vspace*{-0.4cm}
\caption{{\bf Momentum dependence of charged hadron and D meson $R_{AA}$.  } The left panel shows together the experimentally measured 0-5\% central 2.76 Pb+Pb ALICE preliminary $R_{AA}$ data for charged hadrons~\cite{ALICE_h} (the red circles) and D mesons~\cite{ALICE_preliminary} (the blue triangles). The right  panel shows the comparison of the light hadron suppression predictions (the gray band with full-curve boundaries) with the D meson suppression predictions (the gray band with dashed-curve boundaries). Both gray regions correspond to $0.4 < \mu_M/\mu_E < 0.6$, where the upper (lower) boundary on each band corresponds to $\mu_M/\mu_E =  0.6$ ($\mu_M/\mu_E =  0.4$).}
\label{TheoryVsData}
\end{figure*}
 
As discussed above, Figures~\ref{QuarkELossRaa} and~\ref{GLRatio} lead to the expectation that the charged hadron suppression should be significantly larger than the D meson suppression. Surprisingly, this expectation is not confirmed by the experimental data, which are shown in the left panel of Fig.~\ref{TheoryVsData}; these data clearly suggest the same suppression for both pions and D mesons. We can also calculate these $R_{AA}$s through the finite size dynamical energy loss formalism~\cite{MD_LHSupp_2013}, which is shown in the left panel of Fig.~\ref{TheoryVsData}. Even more surprisingly, these calculations are in accordance with the experimental data, i.e. they show the same suppression patterns for light hadrons and D mesons. Moreover, we see that the theoretical predictions even reproduce the experimentally observed smaller suppression of D mesons compared to charged hadrons  in the lower momentum range. This difference is the consequence of the ``dead cone effect''~\cite{Kharzeev}, as can be seen in Fig.~\ref{QuarkELossRaa}. Consequently, we see that our theoretical predictions show a very good agreement with the experimental data, which is in an apparent contradictions with the qualitative expectations discussed above; we will below concentrate on finding the explanation for these unexpected results.

\begin{figure}
\epsfig{file=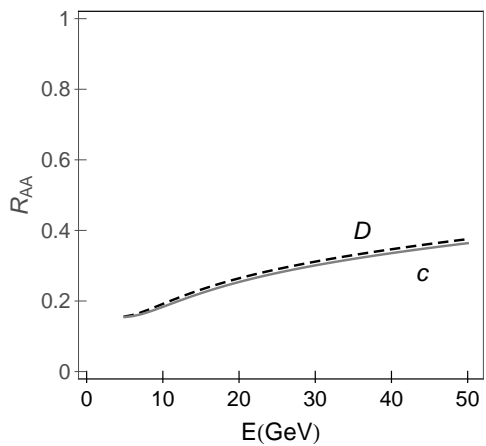,width=2.5in,height=2.3in,clip=5,angle=0}
\vspace*{-0.3cm}
\caption{{\bf Comparison of charm quark and D mason suppression predictions.} The figure shows a comparison of the charm quark suppression predictions (the full curve) with the D meson suppression predictions (the dashed curve), as a function of momentum. Electric to magnetic mass ratio is fixed to $\mu_M/\mu_E =0.5$. }
\label{HeavyFlavorRaa}
\end{figure}

We start by asking how fragmentation functions modify light hadron and D meson suppressions, since these functions define the transfer from the parton to the hadron level. We first analyze how fragmentation functions modify the D mason suppression, compared to the bare charm quark suppression. We see that there is a negligible difference between these two suppression patterns, so that D meson fragmentation does not modify  bare charm quark suppression. Consequently, the D meson suppression is indeed a genuine probe of the charm quark suppression in QCD medium.

\begin{figure*}
\epsfig{file=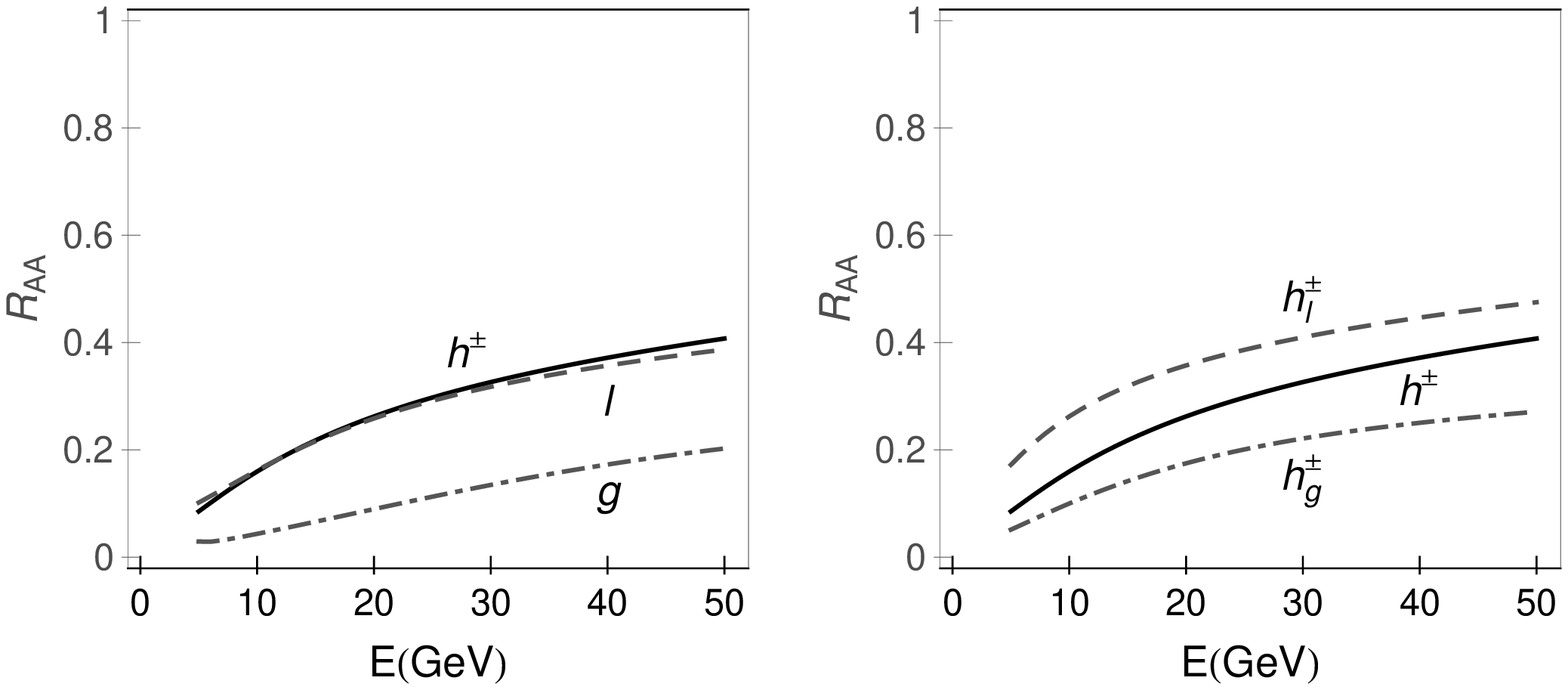,width=5in,height=2.3in,clip=5,angle=0}
\vspace*{-0.4cm}
\caption{{\bf Comparison of the light flavor suppression predictions.} The left panel  shows the comparison of light hadron suppression predictions (the full curve)  with light quark (the dashed curve) and gluon (the dot-dashed curve) suppression predictions, as a function of momentum. On the right panel, the dashed curve shows what would be the light hadron suppression if only light quarks would contribute to light hadrons. The dot-dashed curve shows what would be the light hadron suppression if only gluons would contribute to light hadrons, while the full curve shows the actual hadron suppression predictions. On each panel, electric to magnetic mass ratio is fixed to $\mu_M/\mu_E =0.5$.}
\label{LightFlavorRaa}
\end{figure*}

However, there is a significantly more complex interplay between suppression and fragmentation in charged hadrons. In the left panel of Fig.~\ref{LightFlavorRaa}, we compare the light hadron suppression with the bare light quark and gluon suppression patterns. Surprisingly, we see that the light hadron suppression pattern almost exactly coincides with the bare light quark suppression. This may suggest that gluon jets do not contribute to the light hadron suppression, which is however clearly inconsistent with the significant (even dominant) gluon contribution in light hadrons (see Fig.~\ref{GLRatio}). To further investigate this, in the right panel of Fig.~\ref{LightFlavorRaa}, we show what would be the light hadron suppression if hadrons were composed only by light quark jets (the dashed curve), or only by gluon jets (the dot-dashed curve). We see that, as expected from Fig.~\ref{GLRatio}, the actual light hadron suppression is clearly in between the two above mentioned suppression alternatives, so that both light quarks and gluons indeed significantly contribute to the light hadron suppression. However, by comparing the two panels in Fig.~\ref{LightFlavorRaa}, we see that light hadron fragmentation functions modify the bare light quark and gluon suppressions so that, coincidentally, their ``resultant'' charged hadron suppression  almost identically reproduces the bare light quark suppression. Consequently, the heavy flavor puzzle at LHC is a consequence of a specific combination of the suppression and fragmentation patterns for light partons, and it does not require invoking an assumption of the same energy loss for light partons.

\section{Conclusions}

A major theoretical goal in relativistic heavy ion physics is to develop a theoretical framework that is able to consistently explain both light  and heavy flavor experimental data. We here analyzed the comparison of charged hadron and D meson suppression data in central 2.76 TeV Pb+Pb collisions at LHC, which leads to a surprising puzzle. To analyze this puzzle, we here used a dynamical energy loss formalism. While the solution of this puzzle is inherently quantitative, it can be qualitatively summarized in the following way: Despite the dominant gluon contribution in the charged hadron production, LHC charged hadron suppression turns out to be a genuine probe of bare light quark suppression. The main effect responsible for this key result is the distortion of the bare suppression patterns by the jet fragmentation. Furthermore, the D meson suppression correctly represents charm quark suppression, and bare charm and light quark suppressions are very similar. Taken together, these results in fact explain the observed puzzle, i.e. the fact that light hadron and D meson suppression are measured to be the same  at LHC.

Therefore, the explanation of the puzzle follows directly from pQCD calculations of the energy loss and fragmentation. Furthermore, these calculations directly relate the bare quark suppressions to the experimentally observed charged hadron suppressions. Consequently, the heavy flavor puzzle at LHC is not only a coincidental combination of suppression and fragmentation patterns, but also their serendipitous interplay, which can substantially simplify the interpretation of the relevant experiments.

{\em Acknowledgments:} 
This work is supported by Marie Curie International Reintegration Grant 
within the $7^{th}$ European Community Framework Programme 
(PIRG08-GA-2010-276913) and by the Ministry of Education, Science and Technological 
Development of the Republic of Serbia, under projects No. ON171004 and 
ON173052. I thank I. Vitev and Z. Kang for providing the initial light 
flavor distributions and useful discussions. I also thank M. Cacciari 
for useful discussion on heavy flavor production and decay processes. 
I thank ALICE Collaboration for providing the shown preliminary data, 
and thank M. Stratmann and Z. Kang for help with DSS fragmentation functions.

\end{document}